\newif\ifShowKeys
\ShowKeysfalse

\documentclass[11pt,a4paper]{article} 
\pdfoutput=1
\usepackage[no-natbib-sort]{my-jheppub}

\usepackage{amsmath, amssymb}

\usepackage{mathpazo}
\usepackage{bm}
\usepackage{environ}
\usepackage{mathrsfs}
\usepackage{array,arydshln}

\usepackage{graphicx,epsfig}
\usepackage{epic}
\usepackage{youngtab}
\usepackage{float}
\usepackage{color}
\definecolor{maroon}{rgb}{0.8,0.3,0.}

\usepackage{slashed}
\usepackage[nodayofweek]{date time}

\ifShowKeys \usepackage[notcite]{showkeys} \fi

\usepackage{hyperref}

\usepackage{aurical}
\usepackage[T1]{fontenc}

\allowdisplaybreaks

\newcommand{\be}{\begin{equation}}
\newcommand{\ee}{\end{equation}}

\newcommand{\mc}{\mathcal }

\newcommand{\bv}{\begin{verbatim}}
\newcommand{\ev}{\end{verbatim}}

\newcommand{\la}{\label}

\newcommand{\schr}{Schr\"odinger }

\newcommand{\blue}[1]{\textcolor{blue}{#1}}

\title{On the cusp anomalous dimension in the \\ ladder limit of $\mc N=4$ SYM}
\author[a,b]{Matteo Beccaria} 
\author[a,b]{, Alberto Fachechi} 
\author[a,b]{, Guido Macorini} 

\abstract{
We analyze the cusp anomalous dimension in the (leading) ladder limit of $\mc N=4$ SYM and 
present new results for its higher-order perturbative expansion. We study two different limits with respect 
to the cusp angle $\phi$. The first is the light-like regime where $x = e^{i\,\phi}\to 0$.
This limit is characterised by a non-trivial expansion of the cusp anomaly as a sum of powers of $\log x$,
where the maximum exponent increases with the loop order. 
The coefficients of this expansion have remarkable transcendentality 
features and can be expressed by products of single zeta values.
We show that the whole logarithmic expansion is  fully captured by a solvable Woods-Saxon like 
one-dimensional potential.
From the exact solution, we extract generating functions for the cusp anomaly as well as for the various 
specific transcendental structures appearing therein. The second limit that we discuss is 
the regime of small cusp angle. In this somewhat simpler case, we show how to organise the 
quantum mechanical perturbation theory in a novel efficient way by means of a suitable all-order Ansatz for the ground state of the associated \schr problem.
Our perturbative setup allows to systematically derive  higher-order perturbative corrections in powers of the  
cusp angle as explicit non-perturbative functions of the effective coupling. This series approximation
is compared with the numerical solution of the \schr equation to show that we can achieve very good 
accuracy over the whole range of coupling and cusp angle. Our results have been obtained by relatively simple
techniques. Nevertheless, they provide several  non-trivial tests useful to check the application of 
Quantum Spectral Curve methods to the ladder approximation at non zero $\phi$, in the two limits we studied.

\vfill }

\affiliation[a]{Dipartimento di Matematica e Fisica Ennio De Giorgi,\\
Universit\`a del Salento, Via Arnesano, 73100 Lecce, 
Italy} 

\affiliation[b]{INFN, Via Arnesano, 73100 Lecce, Italy}

\emailAdd{matteo.beccaria@le.infn.it} 
\emailAdd{alberto.fachechi@gmail.com} 
\emailAdd{macorini@nbi.ku.dk} 

\begin{document}



\maketitle
\flushbottom

\section{Introduction}

The study of cusped Wilson loops $W$ was  
initiated by Polyakov in \cite{Polyakov:1980ca} while attempting to view gauge fields as
chiral fields on a loop space. Gauge interactions are interpreted in terms of the  propagation of  infinitely thin rings 
formed by the lines of color-electric flux. The analysis of the quantum properties of  
$W$ led to the introduction of the cusp anomalous dimension 
$\Gamma_{\text{cusp}}(\phi)$ depending on the Euclidean cusp angle $\phi$ and 
appearing in the relation
\be
\la{1.1}
\langle W \rangle \sim e^{-\Gamma_{\text{cusp}}(\phi)\, \log\frac{\Lambda_{\text{UV}}}{\Lambda_{\text{IR}}}},
\ee
where $\Lambda_{\text{UV, IR}}$ are ultraviolet and infrared energy cutoffs. 
In QCD, the cusp anomalous dimension was first 
computed at two loops in \cite{Korchemsky:1987wg,Kidonakis:2009ev} and has been 
recently extended to  three loops in \cite{Grozin:2014hna,Grozin:2015kna}, see also \cite{Kidonakis:2016voy}.
In supersymmetric theories, it is possible to introduce a locally supersymmetric Wilson loop that couples to scalars in addition to gluons. In $\mc N=4$ SYM,  the cusp anomalous dimension has been computed 
at two loops in \cite{Makeenko:2006ds,Drukker:2011za}, at three loops in \cite{Correa:2012nk}, and at four loops
in \cite{Henn:2013wfa}. 

\medskip
The free parameter $\phi$ can be tuned to discuss several interesting  physical regimes. 
At small cusp angle, we are doing perturbation  around the straight line configuration, $\phi=0$, 
that is half  BPS and has no quantum corrections. The first non-trivial term in the small angle expansion
is $\Gamma_{\text{cusp}}(\phi) = -B(\lambda)\, \phi^{2}+\mc O(\phi^{4})$ where $B(\lambda)$ 
is the so-called Bremstrahlung function
 (depending on the planar 't Hooft coupling $\lambda$) which is 
 fully known \cite{Correa:2012at,Fiol:2012sg}.
From $\Gamma_{\text{cusp}}(\phi)$, we can extract the potential for a quark anti-quark pair living 
on a 3-sphere and separated by the angle $\delta=\pi-\phi$. In the limit $\delta\to 0$, the flat space potential 
is recovered. \footnote{The quark anti-quark potential is known at 3 loops at weak coupling
\cite{Erickson:1999qv,Pineda:2007kz,Correa:2012nk,Bykov:2012sc,Stahlhofen:2012zx,Prausa:2013qva,Drukker:2011za}  and at one loop at strong coupling 
\cite{Maldacena:1998im,Rey:1998ik,Forini:2010ek,Chu:2009qt}. It may be treated at all orders by means of the Quantum algebraic curve \cite{Gromov:2016rrp}.}
Finally, one can analytically continue in the cusp angle and consider the limit 
$\varphi = i\,\phi\to \infty$ where it turns out that 
$\Gamma_{\text{cusp}}\sim \varphi\,\Gamma_{\text{cusp}}^{\infty}$. The coefficients 
$\Gamma_{\text{cusp}}^{\infty}$ is 
the anomalous dimension of a null Wilson loop and is 
related to the high-spin behaviour of anomalous dimensions of composite operators 
\cite{Korchemsky:1988si,Korchemsky:1992xv,Alday:2007mf}, governed by the celebrated BES
exact integral equation derived in \cite{Beisert:2006ez} by  integrability methods in $\mc N=4$ SYM.

\medskip
A quite important feature of the $\mc N=4$ SYM case is that it is possible to introduce an additional angle 
in the definition of the Wilson loop \cite{Drukker:1999zq}. The locally supersymmetric Wilson loop 
contains a coupling $\int dt\, \Phi\cdot n\, |\dot x(t)|$, where $\Phi$ is the  vector of 
 the 6 scalars of  $\mc N=4$ SYM and  $x(t)$ is the piece-wise straight quark (anti-quark) trajectory
\cite{Maldacena:1998im}.
The unit vector $n\in S^{5}$ is constant apart from a discontinuous turn at the cusp 
by the angle $\theta$.
A crucial remark made in  \cite{Correa:2012nk} is that the extra parameter $\theta$ may be used to study the 
scaling limit 
\be
\label{1.2}
i\,\theta\to \infty, \qquad \widehat\lambda = \tfrac{1}{4}\,\lambda\,e^{i\,\theta} \quad \text{fixed}.
\ee
The limit (\ref{1.2}) is interesting because it 
selects ladder diagrams and, remarkably, the  cusp anomalous dimension may be identified with 
the ground state energy of a 1d Schr\"odinger equation. In standard notation, it is a function 
$\Gamma^{\text{lad}}(\kappa, \phi) = -\Omega(\kappa, \phi)$ where 
 $\kappa = \widehat\lambda/\pi^{2}$ and $\Omega>0$ is (essentially) the ground state energy 
 of the  \schr problem 
\be
\label{1.3}
\begin{split}
& \bigg(-\frac{d^{2}}{dw^{2}}+V(w, \phi) \bigg)\,\psi(w) = 
-\frac{\Omega^{2}(\kappa, \phi)}{4}\,\psi(w),\qquad w\in \mathbb R,\\
& V(w,\phi) = -\frac{\kappa}{8}\,\frac{1}{\cosh w+\cos\phi}.
\end{split}
\ee
Despite its simplicity, the ladder approximation 
is quite interesting and various remarkable feature of the function $\Gamma^{\rm lad}(\kappa, \phi)$
have been investigated at generic $\phi$ in \cite{Correa:2012nk,Henn:2012qz}. \footnote{
The $\phi\to \pi$ limit is particularly interesting because it allows to extract the flat space quark-antiquark
potential. However, it is difficult because the \schr ground state energy is not analytic in the coupling,  see \cite{Gromov:2016rrp,Beccaria:2016ejo}.}
In this paper, we reconsider its perturbative expansion
in two special limits where we are able to provide new exact results.

\medskip
\noindent{\bf Light-like limit $x\to 0$, where $x=e^{i\,\phi}$.} 

\noindent
As remarked in  \cite{Henn:2012qz}, the limit $x\to 0$ 
is interesting because it connects the velocity-dependent cusp anomalous
dimension  with the light-like cusp anomalous dimension which corresponds to the light-like limit of the edges of the Wilson loop.
From the six loop analysis of \cite{Henn:2012qz}, it is
possible to identify the following remarkable structure \footnote{The expansion (\ref{1.4}) is what is found in the 
ladder approximation. The true cusp anomaly is linear  $\sim\log x$ for $x\to 0$ with remarkable cancellations
deleting the higher powers of $\log x$, see for instance \cite{Henn:2013wfa}. }
\be
\label{1.4}
\begin{split}
\Gamma^{\rm lad} &= \sum_{n=0}^{\infty} \bigg[b^{(n)}(x)\,x^{n}+\mc O(x^{n+1})\bigg]\,(\kappa/4)^{n},
\qquad b^{(n)}(x) = \sum_{k=1}^{2n} b^{(n)}_{k}\,\log^{2n-k}x,
\end{split}
\ee
where $b^{(n)}_{1}$ is rational, $b^{(n)}_{2}=0$, $b^{(n)}_{3}$ is a rational multiple of $\zeta(2)$, 
 $b^{(n)}_{4}$ is a rational multiple of $\zeta(3)$, $b^{(n)}_{5}$ is a rational multiple of $\zeta(4)$,
 and all next coefficients $b^{(n)}_{k}$ are linear combinations of  
 products of simple $\zeta$  values with transcendentality degree $\text{d}=k-1$. 
The degree $\text{d}$ gets a contribution equal to $n$ from $\zeta_{n}$ (for even $n$ the involved transcendental 
constant is $\pi^{2n}$) and is additive with respect to multiplication $\text{d}(AB) = \text{d}(A)+\text{d}(B)$.
 The expansion in (\ref{1.4}) has been determined at six loops by the algorithm discussed in 
\cite{Henn:2012qz} involving harmonic polylogarithms and their small $x$ expansions. In  that 
approach, it is non trivial to explore the above properties of the coefficients in (\ref{1.4}). Here,
we study the $x\to 0$ limit of the ladder
 Schr\"odinger potential by a different analytical approach. 
 In particular, we identify a reduced \schr equation that captures all
 the logarithmic terms in (\ref{1.4}). It is a 1d version of the three-dimensional Woods-Saxon potential. 
 Its ground state is solvable and from its explicit expression we derive several useful generating functions for the 
 the coefficients $b^{(n)}_{k}$. They are exact in $\kappa$ and 
 can be used to systematically obtain long expansions at higher-loop order.

\medskip
\noindent{\bf Small angle $\phi$}

\noindent
For $\phi=0$, the ladder approximation reduces to a \schr equation 
with solvable P\"oschl-Teller potential, see (\ref{1.3}). Perturbation theory in $\phi$ is analytic and takes the form 
\be
\label{1.5}
\Gamma^{\rm lad} = \sum_{n=0}^{\infty} c_{n}(\kappa)\,\phi^{2\,n}, \qquad \kappa = \frac{\widehat \lambda}{\pi^{2}}.
\ee
In \cite{Correa:2012nk}, first order 
Rayleigh-Schr\"odinger perturbation theory has been applied to provide the 
results
\be
\la{1.6}
c_{0}(\kappa) =  \frac{1-\sqrt{1+\kappa}}{2}, \qquad
c_{1}(\kappa) = -\frac{\kappa}{16}\,
\frac{1+\sqrt{1+\kappa}}{1+\kappa+2\,\sqrt{1+\kappa}}.
\ee
The coefficients in the expansion (\ref{1.6}) are interesting because they are 
non-perturbative in the effective 't Hooft coupling 
$\widehat \lambda = \pi^{2}\,\kappa$. We show that it is possible to systematically improve (\ref{1.6}) by implementing 
a perturbation method originally proposed in  \cite{dalgarno1956perturbation}
that typically  works in the case of  polynomial perturbations.
This method has the advantage of 
bypassing the machinery of the Rayleigh-\schr approach.
The resulting algorithm is applied to obtain the coefficient functions $c_{n}(\kappa)$ in closed form for very high 
$n$. The associated long series expansion is successfully compared with the numerical solution of the \schr problem
in the whole range of physical parameters $\kappa, \phi$.

\medskip
The plan of the paper is the following. 
In Sec.~(\ref{sec:light}) we study the light-like limit and provide a master equation that permits to easily  extract 
the whole logarithmic expansion in (\ref{1.4}) at any loop order. In Sec.~(\ref{subsec:generating}), 
we further manipulate the master equation showing how to determine a compact generating function for the various 
transcendentality structures appearing in (\ref{1.4}). 
In Sec.~(\ref{sec:small}), we treat the small $\phi$ perturbative expansion of the
cusp anomalous dimension. The higher order results are checked at large $\kappa$ in Sec.~(\ref{subsec:large}).
In Sec.~(\ref{subsec:num}), we show that our expansions can be used to provide 
the correct cusp anomaly at all $\kappa$ and $\phi$ with great accuracy.
Various appendices collect long results and 
related discussions.

\section{The light-like limit $x\to 0$}
\label{sec:light}

\begin{figure}[h]
\centering
\includegraphics[width=0.8\textwidth]{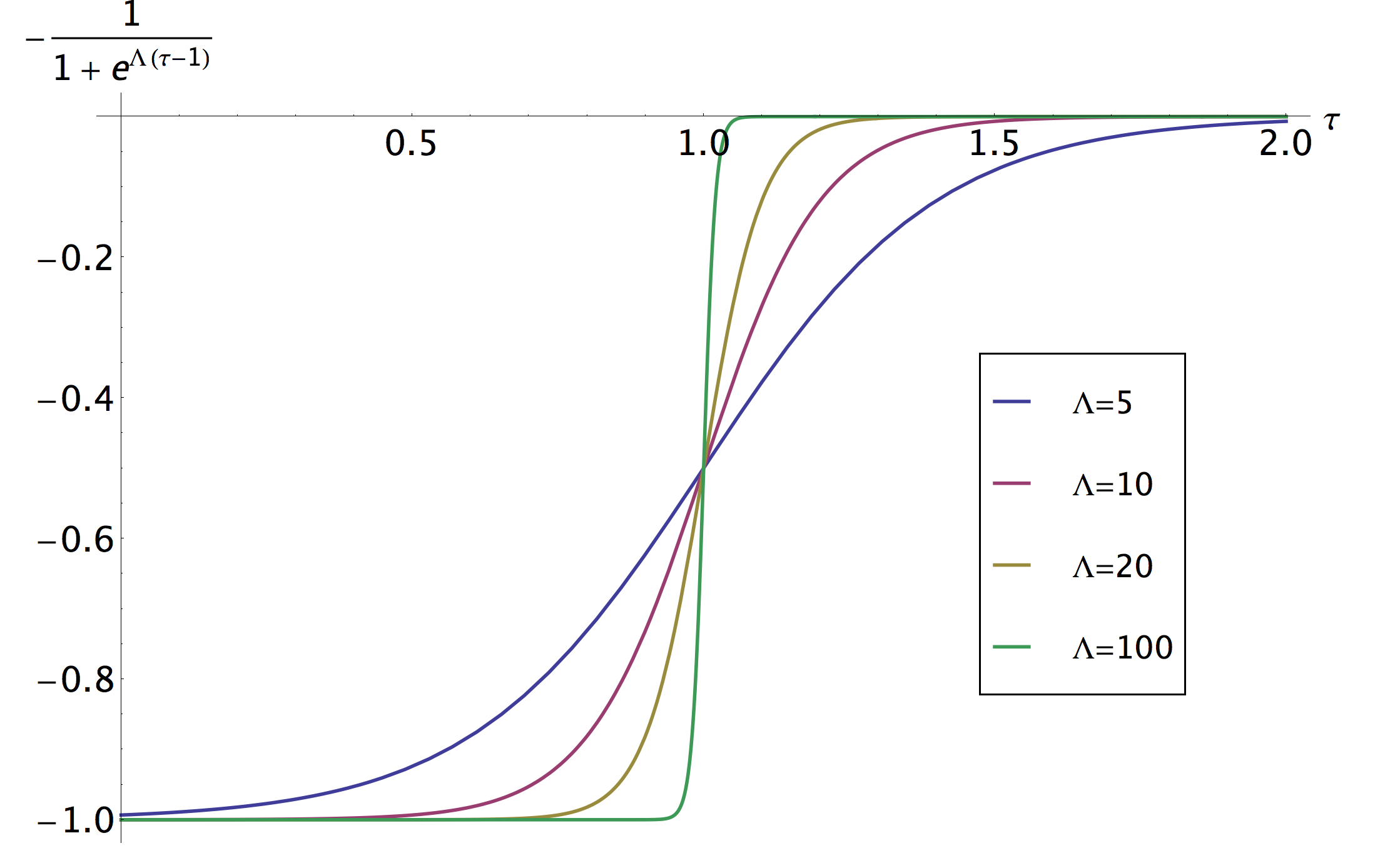}
\caption{Plot of the approximate potential $\frac{1}{1+ e^{\Lambda(\tau-1)}}$.
As $\Lambda = -\log x$ is increased, the potential approaches a finite depth square well.}
\label{fig:1}
\end{figure}

As we discussed in the introduction, the first limit we want to treat is $x = e^{i\phi}\to 0$.
To explain what we are going to compute, it is 
convenient to recall the results of \cite{Henn:2012qz} providing the weak-coupling expansion (\ref{1.4})
at 6-loops.
We give it in terms of the coefficients $\Omega_{n}(x)$ appearing in, see their Eq.~(3.40),
\be
\la{2.1}
\Gamma^{\text{lad}} = -\sum_{n=1}^{\infty}
\bigg(\frac{1}{4}\,
\frac{\kappa\,x}{1-x^{2}}
\bigg)^{n}
\Omega_{n}(x).
\ee
When $x\to 0$, the coefficient functions $\Omega_{n}(x)$ have the structure outlined 
in (\ref{1.4}). Writing only the first four 
non vanishing leading terms, the six loop results obtained in \cite{Henn:2012qz} are \footnote{The exact expression at 
two loops is quite simple and reads
\be
\notag 
\Omega_{2}(x) = -4\, \text{Li}_3(x^2)+4\, \text{Li}_2(x^2) \log x+\frac{4}{3}\, \log^3 x
+\frac{2}{3} \pi ^2 \log x+4 \zeta_{3}.
\ee
}
\begin{align}
\la{2.2}
\Omega_{1}(x) &= -2\,\log x, \notag\\
\Omega_{2}(x) &= \frac{4}{3}\,\log^{3}x+4\,\zeta_{2}\,\log x+4\,\zeta_{3}+\mc O(x), \notag\\
\Omega_{3}(x) &= -\frac{8}{5}\,\log^{5}x-\frac{32}{3}\,\zeta_{2}\,\log^{3}x-8\,\zeta_{3}\,\log^{2}x
-24\,\zeta_{4}\,\log x+\mc O(1), \notag\\
\Omega_{4}(x) &= \frac{736}{315}\,\log^{7}x+\frac{368}{15}\,\zeta_{2}\,\log^{5}x
+16\,\zeta_{3}\,\log^{4}x+152\,\zeta_{4}\,\log^{3}x+\mc O(\log^{2}x), \\
\Omega_{5}(x) &= -\frac{2144}{567}\,\log^{9}x-\frac{17152}{315}\,\zeta_{2}\,\log^{7}x
-\frac{1472}{45}\,\zeta_{3}\,\log^{6}x-\frac{2896}{5}\,\zeta_{4}\,\log^{5}x+\mc O(\log^{4}x), \notag\\
\Omega_{6}(x) &= \frac{339008}{51975}\,\log^{11}x+\frac{339008}{2835}\,\zeta_{2}\,\log^{9}x
+\frac{4288}{63}\,\zeta_{3}\,\log^{8}x+\frac{12800}{7}\,\zeta_{4}\,\log^{7}x+\mc O(\log^{6}x).\notag
\end{align}
The omitted terms have a uniform transcendentality as discussed in the introduction. \footnote{
The degree 4 terms in (\ref{2.2}) are proportional to $\pi^{4}$. This is written as a rational multiple of $\zeta_{4}$, but
of course one may also use $\zeta_{2}^{2}$. }
They can be found in 
\cite{Henn:2012qz} up to 6-loops. Just to give an example, the complete expression of $\Omega_{6}$ is 
\begin{align}
\label{2.3}
&\Omega_6(x) \,=\, 
\frac{339008}{51975} \, \log^{11} x + \frac{339008}{2835}\,\zeta_{2}\, \log^9 x+
\frac{4288}{63} \,\zeta_{3}\, \log^8 x+\frac{1280 \,\pi ^4}{63}\, \log^7 x 
 \notag \\&\quad
+\bigg( \frac{17152}{135} \,\pi ^2 \, \zeta _3 +\frac{10688 \, \zeta _5}{45}\bigg) \, \log^6 x+
\bigg(\frac{2944}{15} \, \zeta _3^2 +\frac{110944 \,\pi ^6}{14175}\bigg)\, \log^5 x
 \notag \\&\quad
+\bigg(\frac{2896}{45} \,\pi ^4 \, \zeta _3 +\frac{688}{3} \,\pi^2 \, \zeta _5 +
528 \, \zeta _7\bigg) \, \log^4 x+\bigg(\frac{1472}{9} \,\pi ^2 \, \zeta _3^2
  \\&\quad
+\frac{2432}{3} \, \zeta _3 \, \zeta _5 
+\frac{40024 \,\pi ^8}{42525}\bigg)\, \log^3 x+\bigg(
128 \, \zeta _3^3 +\frac{2528}{315} \,\pi ^6 \, \zeta _3
 \notag \\&\quad
+\frac{632}{15} \,\pi ^4 \, \zeta _5 +192 \,\pi ^2 \, \zeta _7 +664\, \zeta _9\bigg) \, \log^2 x+
\bigg(\frac{304}{15} \,\pi ^4 \, \zeta _3^2 
 \notag \\&\quad
+336 \, \zeta _5^2+192 \,\pi ^2 \, \zeta _3 \, \zeta _5 +672 \, \zeta _3 \, \zeta _7+\frac{2764 \,\pi ^{10}}{155925}\bigg)\,\log x+
\frac{128}{9} \,\pi ^2 \, \zeta _3^3
 \notag \\&\quad
+\frac{248 \,\pi ^8 \, \zeta _3}{2835}+
160\, \zeta _3^2 \, \zeta _5 +\frac{68 \,\pi ^6 \, \zeta _5}{105}+
\frac{24 \,\pi ^4\, \zeta _7}{5}+\frac{340 \,\pi ^2 \, \zeta _9}{9}+
372 \, \zeta_{11}.
\end{align}
An algorithm to compute the full $x$-dependence of $\Omega_{n}$ has been proposed in \cite{Henn:2012qz}
and is based on a recursive representation in terms of harmonic polylogarithms. 
The light-like limit can be treated as a special case or by a simplification of the algorithm. The extension to higher loops is certainly possible, but quite involved. In  \cite{Henn:2012qz}, it has been remarked that only 
powers of single zeta values appear in the asymptotic expansion, at least up to six loops. Here, we show how to generate expansions like (\ref{2.3}) in a 
simple way. We want to 
select the logarithmic terms in (\ref{2.2}) and neglect $\mc O(x)$ corrections. To this aim,  
it is convenient to scale the independent
variable in the Schr\"odinger equation (\ref{1.3}) and introduce $\tau$ by setting $w = \Lambda\,\tau$, where
$\Lambda = -\log x$ is a parameter that will be sent to $+\infty$. The potential 
becomes, for $\tau\ge 0$,
\be
\la{2.5}
\begin{split}
V &= -\frac{\kappa}{8}\,\frac{1}{\cosh(\tau\, \log x)+\frac{1}{2}(x+x^{-1})} = 
-\frac{\kappa\,x}{4}\,\frac{1}{1+e^{\Lambda\,(\tau-1)}+e^{-\Lambda\,(\tau+1)}+e^{-2\,\Lambda}}\\
&=-\frac{\kappa\,x}{4}\,\frac{1}{1+e^{\Lambda\,(\tau-1)}}+\mc O(x).
\end{split}
\ee
For $\tau<0$, the potential in (\ref{2.5}) is extended by $\tau\to -\tau$ symmetry. In the following, we shall restrict to the region $\tau\ge 0$. The potential in (\ref{2.5}) is a one-dimensional version of the 
Woods-Saxon confining model. When $\Lambda\to +\infty$,
we have a negative constant $-\tfrac{1}{4}\,\kappa\, x$ for $0\le \tau< 1$ and zero for $\tau>1$, see Fig.~(\ref{fig:1}). 
The \schr equation with the potential (\ref{2.5}) can be solved exactly with boundary conditions 
\be
\la{2.6}
\psi'(0) = 0, \qquad \psi(+\infty)=0.
\ee
The solution vanishing at infinity is (up to a complex normalization constant) 
\be
\la{2.7}
\psi(\tau) = y^{\frac{i}{2}\,\beta}\,\frac{(\Omega+i\,\beta)\,\Gamma(\frac{\Omega+i\,\beta}{2})^{2}}
{\Gamma(1+i\,\beta)}\,
{}_{2}F_{1}\bigg(\frac{-\Omega+i\,\beta}{2}, \frac{\Omega+i\,\beta}{2}, 1+i\,\beta; -y\bigg)-\text{c.c}.
\ee
where
\be
\la{2.8}
y = x\, e^{\Lambda\, \tau}, \qquad \beta = \sqrt{\kappa \, x-\Omega^{2}}.
\ee
Imposing the second boundary condition $\psi'(0)=0$ and neglecting all terms that vanish as $x\to 0$
faster than any power of $\log x$, we arrive at the master equation 
\be
\la{2.9}\boxed{
2\,\cos(\tfrac{1}{2}\,\Lambda\,\beta) = \frac{\beta+i\,\Omega}{\beta-i\,\Omega}\,
\frac{\Gamma(1+i\,\beta)}{\Gamma(1-i\,\beta)}\,\bigg[
\frac{\Gamma(\frac{\Omega-i\,\beta}{2})}{\Gamma(\frac{\Omega+i\,\beta}{2})}
\bigg]^{2}+
\frac{\beta-i\,\Omega}{\beta+i\,\Omega}\,
\frac{\Gamma(1-i\,\beta)}{\Gamma(1+i\,\beta)}\,\bigg[
\frac{\Gamma(\frac{\Omega+i\,\beta}{2})}{\Gamma(\frac{\Omega-i\,\beta}{2})}
\bigg]^{2}.
}
\ee
Expansion of (\ref{2.9}) is quite simple. One simply writes $\Omega$ as a power series 
in $\kappa$ (actually $\kappa\,x$) starting at order $\mc O(\kappa)$, and uses the definition of $\beta$ in (\ref{2.8}).
This procedure fully reproduces the six loop results in (\ref{2.2}), and can be extended at higher orders. For instance, 
at seven loops, we find the new expression
\begin{align}
\label{2.10}
 &\Omega_7(x) \,=\, -
\frac{2041856}{173745}\, \log^{13} x -\frac{4083712 \,\pi ^2}{93555} \, \log^{11} x-
\frac{678016}{4725}\,\zeta_{3} \, \log^{10} x 
\notag  \\&\quad
 - \frac{2482048 \,\pi ^4}{42525}\, \log^9 x
- \bigg( \frac{339008}{945} \,\pi ^2 \, \zeta _3 +\frac{170176}{315}\,\zeta_{5}\bigg) \, \log^8 x
-\bigg(\frac{34304}{63} \, \zeta _3^2 
 \notag \\&\quad
+\frac{10054144\,\pi ^6}{297675}\bigg) \, \log^7 x 
-\bigg(\frac{2560}{9} \,\pi ^4 \, \zeta _3+\frac{4096}{5} \,\pi ^2 \, \zeta _5 +
\frac{4288 \, \zeta _7}{3}\bigg)\, \log^6 x
 \notag \\&\quad
-\bigg(\frac{34304}{45} \,\pi ^2 \, \zeta _3^2 +\frac{42752}{15}\, \zeta _3 \, \zeta _5 +\frac{341456 \,\pi ^8}{42525}\bigg)\, \log^5 x -\bigg(
\frac{5888}{9} \, \zeta _3^3 
 \notag \\&\quad
+\frac{221888 \,\pi ^6\, \zeta _3}{2835} +\frac{14768}{45} \,\pi ^4 \, \zeta _5+\frac{3344}{3} \,\pi ^2 \, \zeta _7
+
\frac{22384 \, \zeta _9}{9}\bigg)\, \log^4 x
 \notag \\&\quad
-\bigg(\frac{11584}{45} \,\pi ^4 \, \zeta _3^2 + 2112 \, \zeta _5^2+\frac{5504}{3} \,\pi ^2 \, \zeta _3 \, \zeta _5+
4224 \, \zeta _3\, \zeta _7 
 \notag \\&\quad
+\frac{95008 \,\pi ^{10}}{155925} \bigg)\, \log^3 x
-\bigg( \frac{2944}{9} \,\pi^2 \, \zeta _3^3+
\frac{80048 \,\pi ^8 \, \zeta _3 }{14175}+2432\, \zeta _3^2 \, \zeta _5 
 \notag \\&\quad
+\frac{4256}{135} \,\pi ^6 \, \zeta _5+\frac{808}{5} \,\pi ^4 \, \zeta _7 +
\frac{6656}{9} \,\pi ^2 \, \zeta_9 +2584 \, \zeta _{11}\bigg) \, \log^2 x
  \\&\quad
-\bigg(128 \, \zeta _3^4 +\frac{5056}{315} \,\pi^6 \, \zeta _3^2 +384 \,\pi ^2 \, \zeta _5^2 +
\frac{2528}{15} \,\pi ^4\, \zeta _3 \, \zeta _5
 \notag \\&\quad
+768 \,\pi ^2 \, \zeta _3 \, \zeta _7+2592 \, \zeta _5\, \zeta _7+2656 \, \zeta _3 \, \zeta _9 +\frac{43688 \,\pi ^{12}}{6081075}\bigg)\, \log x 
-\frac{608}{45} \,\pi ^4 \, \zeta _3^3
 \notag \\&\quad
-672 \, \zeta _3 \, \zeta _5^2-
\frac{5528\,\pi ^{10} \, \zeta _3}{155925}-192 \,\pi ^2 \, \zeta _3^2 \, \zeta_5
-\frac{248 \,\pi ^8 \, \zeta _5}{945}-672 \, \zeta _3^2 \, \zeta_7-\frac{68 \,\pi ^6 \, \zeta _7}{35}
 \notag \\&\quad
-
\frac{136 \,\pi ^4 \, \zeta _9}{9}-124\,\pi ^2 \, \zeta _{11}-1260 \, \zeta _{13}
+\mc O(x) .\notag
\end{align}
The similar 8-loop result is collected in App.~(\ref{app:eight}).

\subsection{Generating functions and transcendentality expansion }
\label{subsec:generating}

Given the plain structure in (\ref{2.2}) and (\ref{2.10}), it is tempting to understand what is the generating
function for the various coefficients $b^{(n)}_{k}$, see (\ref{1.4}). This may be achieved by means of the
 following trick. The $\Gamma$
functions in the r.h.s.
of (\ref{2.9}) are the only source of transcendental contributions. We can consider the r.h.s. of (\ref{2.9}) 
with fixed ratio 
$\beta/\Omega$ and expand at small $\beta$ the resulting expression. 
This expansion reads
\begin{align}
\la{2.11}
& \text{ r.h.s. of (\ref{2.9})} = 2\,\frac{\beta ^2-\Omega ^2}{\beta ^2+\Omega ^2}-\frac{2\, \beta^2\, \Omega
   ^2}{3\, (\beta ^2+\Omega ^2)}\,\bm{\pi^{2}}+2\, \beta ^2\, \Omega\, \, \bm{\zeta_{3}}\\
   &+\frac{\beta ^2\,
   \Omega ^2\,  (\Omega ^2-\beta ^2)}{60\, (\beta ^2+\Omega ^2)}\,\bm{\pi^{4}}+\frac{1}{6}\, \beta
   ^2 \Omega\,\bigg[  \bm{\pi ^2\, \zeta_{3}} (\beta ^2-\Omega ^2)+3\, \bm{\zeta_{5}}\, (\Omega ^2-3\,
   \beta ^2)\bigg]+\notag\\
   &+\frac{\beta^{2}}{7560\,(\beta^{2}+\Omega^{2})}\,\bigg[
   \bm{\pi^{6}}\,\Omega^{2}\,(5\beta^{4}+16\beta^{2}\Omega^{2}+5\Omega^{4})
   -1890\,\bm{\zeta_{3}^{2}}\,(\beta^{2}-\Omega^{2})\,(\beta^{2}+\Omega^{2})^{2}
   \bigg]\notag \\
   & -\frac{\beta^{2}\Omega}{360}\,\bigg[
   -45\,\bm{\zeta_{7}}\,(\Omega^{2}-3\beta^{2})^{2}+15\,\bm{\pi^{2}\zeta_{5}}\,(3\beta^{4}-4\beta^{2}\Omega^{2}
   +\Omega^{4})\notag\\
   &\qquad \qquad +\bm{\pi^{4}\zeta_{3}}\,(\beta^{4}+8\beta^{2}\Omega^{2}+\Omega^{4})
   \bigg]+\cdots, \notag
\end{align}
and additional contributions may be obtained with no problems. In (\ref{2.11}), we have  written in bold face the transcendental constants.
The terms in (\ref{2.11}) are naturally ordered by 
 increasing transcendentality degree (we remind that $\pi^{2n}$ and $\zeta_{n}$ contribute $n$ units). This means that 
we can decompose $\Omega$ as a sum of contributions of increasing degree and solve (\ref{2.9})
for each of them. The first term in (\ref{2.11}) has degree zero and  gives the leading order condition 
\be
\la{2.12}
\sqrt{\kappa\,x-\omega^{2}_{0}}\,\tan\bigg(\frac{\log x}{2}\,\sqrt{\kappa\,x-\omega^{2}_{0}}\bigg)+\omega_{0}=0.
\ee
Expanding (\ref{2.12}) at small $x$ gives
\be
\la{2.13}
\begin{split}
\omega_{0} &= -2\,x\,\log x\,\frac{\kappa}{4}+\frac{4}{3}\,x^{2}\,\log^{3}x\,\bigg(\frac{\kappa}{4}\bigg)^{2}
-\frac{8}{5}\,x^{3}\,\log^{5}x\,\bigg(\frac{\kappa}{4}\bigg)^{3}
+\frac{736}{315}\,x^{4}\,\log^{7}x\,\bigg(\frac{\kappa}{4}\bigg)^{4}\\
& -\frac{2144}{567}\,x^{5}\,\log^{9}x\,\bigg(\frac{\kappa}{4}\bigg)^{5}
+\frac{339008}{51975}\,x^{6}\,\log^{11}x\,\bigg(\frac{\kappa}{4}\bigg)^{6}+\cdots.
\end{split}
\ee
Comparing (\ref{2.13}) with (\ref{2.2}), we see that the compact relation (\ref{2.12}) captures all the leading logarithms for $x\to 0$ and, of course, may be extended at arbitrarily higher order with minor effort. 
A simple way to do this is to notice that (\ref{2.12}) implies the following differential constraint for 
$\omega_{0} = f(\kappa\,x)$ with 
\be
\la{2.14}
2\,\text{k}\,\bigg[\log x\,f(\text{k})-2\bigg]\,f'(\text{k})+2\,f(\text{k})-\log x\, \text{k} = 0.
\ee
Starting with $f(\text{k}) = -1/2\,\log x\, \text{k}+\cdots$, one gets from (\ref{2.14}) a simple recursion for the coefficients in (\ref{2.13}). The convergence properties of the expansion (\ref{2.13}) are discussed in App.~(\ref{app:conv}). To illustrate what happens beyond the leading (rational) logarithms, we present 
the contributions with transcendentality up to 5, {\em i.e.} proportional to 
$\pi^{2}, \zeta_{3}$,  $\pi^{4}$,  $\zeta_{5}$, and $\pi^{2}\,\zeta_{3}$ as an illustrative
example. After some some straightforward manipulations, we get from (\ref{2.9}) using (\ref{2.11})
\begin{align}
\la{2.15}
\Omega &= \omega_{0}+\bm{\zeta_{2}}\,\frac{\omega_{0}\,(\kappa\,x-\omega^{2}_{0})}{\omega_{0}\,\log x-2}
-\bm{\zeta_{3}}\,\frac{\kappa\,x\,(\kappa\,x-\omega^{2}_{0})}{2\,(\omega_{0}\,\log x-2)} -\bm{\zeta_{4}}\,\frac{\omega_{0}\,(\kappa\,x-\omega_{0}^{2})}{4\,(\omega_{0}\,\log x-2)^{3}}\times \\
& \bigg[
\log^{2}x\,\omega_{0}^{2}\,(\kappa\,x-2\,\omega_{0}^{2})-\log x\,\omega_{0}\,(9\,\kappa\,x-28\,\omega_{0}^{2})
+2\,(12\,\kappa\,x-29\,\omega_{0}^{2})
\bigg] \notag \\
& +\bm{\zeta_{5}}\,\frac{\kappa\,x\,(3\,\kappa\,x-4\,\omega_{0}^{2})\,(\kappa\,x-\omega_{0}^{2})}{8\,(\omega_{0}\,\log x-2)}+\bm{\pi^{2}\,\zeta_{3}}\,\frac{\kappa\,x\,(\kappa\,x-\omega_{0}^{2})
(2\,\kappa\,x+3\,\log x\,\omega_{0}^{3}-8\,\omega_{0}^{2})}{12\,(\omega_{0}\,\log x-2)^{3}}
+\cdots\ , \notag
\end{align}
where again we have written in bold face the transcendental constants. Dots in (\ref{2.15}) stand for terms
with transcendentality $\text{d}>5$.
This is a compact generating function for the considered terms. If we plug (\ref{2.13}) into (\ref{2.15}), we recover the 
associated contributions in (\ref{2.2}). For instance, one obtains immediately the following expansions
valid up to 14 loops (recall that $\Omega_{7}$ and $\Omega_{8}$ are given in (\ref{2.10}) and (\ref{A.1}) respectively)
\begin{align}
\la{2.16}
\Omega_9 \,=\,& -\frac{3863148032}{92775375} \, \log^{17}x -\frac{61810368512}{49116375} \, \zeta _2 \, \log^{15}x-\frac{27910995968}{42567525} \, \zeta _3 \, \log^{14}
\notag \\&
-\frac{8434823168}{225225} \, \zeta _4 \, \log^{13}  -\left(\frac{3488874496}{1403325}\pi ^2 \zeta_3 + \frac{1265275904}{467775} \zeta_5\right) \log^{12}x + \dots\notag \\
\Omega_{10} \,=\,& \frac{30013777507328}{371231385525} \, \log^{19}x+\frac{30013777507328}{10854718875} \, \zeta _2 \, \log^{17}x+\frac{7726296064}{5457375} \, \zeta _3 \, \log^{16}x
\notag \\&
+\frac{1569005477888}{16372125} \, \zeta _4 \, \log^{15}x + \left(  \frac{61810368512}{9823275} \,\pi ^2 \zeta _3 + \frac{36584833024}{6081075} \zeta _5 \right) \log^{14}x+\dots,\notag \\
\Omega_{11} \,=\,& -\frac{2824240881547264 }{17717861581875}\, \log^{21}x-\frac{11296963526189056}{1856156927625} \, \zeta _2 \, \log^{19}x
\notag  \\&
-\frac{60027555014656}{19538493975} \, \zeta _3 \, \log^{18}x-\frac{871572721432576}{3618239625} \, \zeta _4 \, \log^{17}x
\notag  \\&
-\left(\frac{30013777507328}{1915538625} \pi ^2 \zeta _3 +\frac{8533890092032}{638512875} \zeta _5\right) \log^{16}x +\dots, \notag \\
\Omega_{12} \,=\,& \frac{747290210255691776}{2348038513445625} \, \log^{23}+\frac{373645105127845888}{27842353914375} \, \zeta _2 \, \log^{21}x
\\&
+\frac{5648481763094528}{843707694375} \, \zeta _3 \, \log^{20}x+\frac{123198569973917696}{206239658625} \, \zeta _4 \, \log^{19}x
\notag  \\&
+ \left( \frac{11296963526189056 }{293077409625}\pi ^2 \zeta _3 +\frac{170578599356416}{5746615875}  \zeta _5 \right)\, \log^{18}x\dots ,\notag \\
\Omega_{13} \,=\,& -\frac{182689999846621413376}{284473896821296875} \, \log^{25}x-\frac{1461519998772971307008}{49308808782358125} \, \zeta _2 \, \log^{23}x
\notag \\&
-\frac{1494580420511383552}{102088631019375} \, \zeta _3 \, \log^{22}x-\frac{95233730539048824832}{64965492466875} \, \zeta _4 \, \log^{21}x
\notag  \\&
-\left(\frac{373645105127845888 }{3977479130625}\pi ^2 \zeta _3 +\frac{611963637598932992}{9280784638125}  \zeta _5 \right)\, \log^{20}x+ \dots,\notag \\
\Omega_{14}\,=\,& \frac{242473345197975650369536}{185436341599368234375} \, \log^{27}x+\frac{242473345197975650369536}{3698160658676859375} \, \zeta _2 \, \log^{25}x
\notag \\&
+\frac{365379999693242826752}{11378955872851875} \, \zeta _3 \, \log^{24}x+\frac{58635553926660888199168}{16436269594119375}  \, \zeta _4 \, \log^{23}
\notag  \\&
+ \left( \frac{1461519998772971307008 }{6431583754220625}\pi ^2 \zeta _3 +\frac{62835830527435194368}{428772250281375} \zeta _5 \right)\log^{22}x+\dots.\notag
\end{align}
and so on. Having  long series of this kind allows to recognize some interesting pattern. For instance, the 
ratio of the NLO logarithm coefficient (proportional to $\zeta_{2}$) 
to the coefficient of the LO logarithm is simply, see (\ref{1.4})
\be
\la{2.17}
\frac{b^{(n)}_{3}}{b^{(n)}_{1}} = \frac{2\,(n-1)(2n-1)}{n}\,\zeta_{2},
\ee
where $n$ is the loop order. This relation can be proved rigorously starting from (\ref{2.14}) and converting the 
factors in (\ref{2.17}) into differential operators $n\to \kappa \partial_{\kappa}$. Another simple relation concerns
the ratio of the NNLO logarithm coefficient (proportional to $\zeta_{3}$), 
\be
\la{2.18}
\frac{b_{4}^{(n)}}{b_{1}^{(n-1)}} = 2\,(3-2\,n)\,\zeta_{3},
\ee
where the shift in the index of $b_{1}$ is a non trivial fact.
The vanishing $b_{2}^{(n)}=0$ and the 
uniform transcendentality property of the expansion (\ref{1.4}) are thus   direct consequences of the 
relation (\ref{2.11}) since all coefficients in (\ref{2.13}) are rational, given (\ref{2.14}).
As a final remark, we notice that 
the leading order equation (\ref{2.12}) is familiar from the discussion of elementary quantum mechanics in a 
one-dimensional finite well. This is not accidental and the relation is fully spelled out in App.~(\ref{app:qm}).

\section{Higher-order expansion at small $\phi$}
\label{sec:small}

To study the problem (\ref{1.3}) at small $\phi$, we begin by setting \footnote{
The invertibility of (\ref{3.1}) is not an issue here. Our aim is to write the wave function in terms of the variable $y$
that will capture in a simple way the dependence on $w$. Boundary conditions are obvious from (\ref{1.3}).}
\be
\la{3.1}
y = \frac{1}{\cosh w+1}
\ee
The expansion of the potential is polynomial in $y$
\be
\la{3.2}
\frac{1}{\cosh w+\cos\phi} = y+\frac{y^{2}}{2}\,\phi^{2}+\bigg(
\frac{y^{3}}{4}-\frac{y^{2}}{24}\bigg)\,\phi^{4}+
\bigg(\frac{y^{4}}{8}-\frac{y^{3}}{24}+\frac{y^{2}}{720}\bigg)\,\phi^{6}+\cdots.
\ee
The kinetic term is also simple
\be
\la{3.3}
\psi''(w) = y^{2}\,(1-2\,y)\,\psi''(y)+y\,(1-3\,y)\,\psi'(y).
\ee
The exact  ground state wavefunction at $\phi=0$ is known and reads
\be
\la{3.4}
\psi_{0}(y) = y^{\Omega/2}, \qquad \text{with}\qquad \kappa = 4\,\Omega\,(\Omega+1).
\ee
We now make the educated perturbative Ansatz
\be
\la{3.5}
\begin{split}
\psi_{0}(y, \phi) &= y^{\Omega/2}\, \bigg(
1+\sum_{n=1}^{\infty}f_{n}(y)\,\phi^{2n}\bigg), \qquad 
\kappa = 4\,\Omega\,(\Omega+1)+\sum_{n=1}^{\infty} \delta\kappa_{n}(\Omega)\,\phi^{2n},
\end{split}
\ee
where  $\delta\kappa_{n}(\Omega)$
are functions of the ground state energy $\Omega$. 

\medskip
We emphasize that it is crucial to set up the perturbative 
expansion according to (\ref{3.5}), {\em i.e.} expanding $\kappa$, 
instead of writing a more natural perturbative expansion of $\Omega$ as a function of $\kappa$. This has the advantage 
that the unperturbed ground state in (\ref{3.4}) -- appearing as a factor in (\ref{3.5}) -- is not changed during the procedure. Replacing (\ref{3.5})
into the Schr\"odinger equation, we get
\begin{align}
\la{3.6}
& y^{-\Omega}\,\sqrt{1-2\,y}\,\frac{d}{dy} \bigg[y^{\Omega+1}\,\sqrt{1-2y}\,f_{1}'(y)\bigg]
 +\frac{y}{4}\,\Omega\,(\Omega+1)+\frac{1}{8}\,\delta\kappa_{1} = 0, \\
& y^{-\Omega}\,\sqrt{1-2\,y}\,\frac{d}{dy} \bigg[y^{\Omega+1}\,\sqrt{1-2y}\,f_{2}'(y)\bigg]
 +\frac{y}{4}\,\Omega\,(\Omega+1)
 +f_{1}(y)\,\bigg(\frac{\delta\kappa_{1}}{8}+\frac{y}{4}\,\Omega\,(\Omega+1)\bigg)\notag \\
 &\qquad +
 \frac{1}{48}\,y\,(6y-1)\,\Omega\,(\Omega+1)+\frac{1}{8}\delta\kappa_{2}+\frac{y}{16}\,\delta\kappa_{1}
  = 0, \notag \\
  & \qquad\qquad \cdots \notag
\end{align}
and so on. Clearly, the  equations in the chain  (\ref{3.6}) can be integrated one after the other. Imposing 
boundary conditions, we fix the coefficients $\delta\kappa_{n}$. Actually, inspection of the results shows that 
the general form of the corrected wave-function is 
\be
\la{3.7}
\boxed{
\begin{split}
\psi_{0}(y, \phi) &= y^{\Omega/2}\, \bigg(
1+\sum_{n=1}^{\infty}\phi^{2n}\, \sum_{k=1}^{n} a_{n,k}\,y^{k}
\bigg).
\end{split}
}
\ee
In other words, the corrections $f_{n}(y)$ are simple polynomials in $y$ ! This remarkable feature is recurrent in 
quantum mechanical problems with a perturbation in the form of a polynomial, see the original proposal in \cite{dalgarno1956perturbation}
or, for instance, \cite{fernandez2000introduction}.
The explicit solution for the first three non trivial corrections is 
\begin{align}
\label{3.8}
f_1(y) = &y \frac{\,\Omega  (\,\Omega +1)}{8 \,\Omega +12},
\notag \\
f_2(y) = &y^2  \frac{\,\Omega  (\,\Omega +1) (\,\Omega +2) (\,\Omega +3)}{32 (2 \,\Omega +3) (2 \,\Omega +5)}- y \frac{\,\Omega  (\,\Omega +1) \left(8 \,\Omega ^3+41 \,\Omega ^2+63 \,\Omega +27\right)}{48 (2 \,\Omega +3)^3 (2 \,\Omega +5)},
 \\
f_3(y) = & y^3  \frac{\,\Omega  (\,\Omega +1) (\,\Omega +2) (\,\Omega +3) (\,\Omega +4) (\,\Omega +5)}{384 (2 \,\Omega +3) (2 \,\Omega +5) (2 \,\Omega +7)}-
\notag \\ & \qquad
y^2
\frac{\,\Omega  (\,\Omega +1) (\,\Omega +2) (\,\Omega +3) \left(8 \,\Omega ^3+49 \,\Omega ^2+87 \,\Omega +45\right)}{192 (2 \,\Omega +3)^3 (2 \,\Omega +5) (2 \,\Omega +7)}+
\notag \\ & \qquad
y
\frac{\,\Omega
  (\,\Omega +1) \left(64 \,\Omega ^6+738 \,\Omega ^5+3383 \,\Omega ^4+7803 \,\Omega ^3+9378 \,\Omega ^2+5454 \,\Omega +1215\right)}{1440 (2 \,\Omega +3)^5 (2 \,\Omega +5) (2 \,\Omega +7)}.
\end{align}
and
\begin{align}
\la{3.10}
\delta \kappa_1  = & -\frac{2\, \Omega  (\,\Omega +1)^2}{2 \,\Omega +3}, \notag \\
\delta \kappa_2  = & \frac{\,\Omega  (\,\Omega +1)^2 \left(8 \,\Omega ^3+41 \,\Omega ^2+63\, \Omega +27\right)}{6 (2 \,\Omega +3)^3 (2 \,\Omega +5)},\\
\delta \kappa_3  = & -\frac{\,\Omega  (\,\Omega +1)^2 \left(64\, \Omega ^6+738\,\Omega ^5+3383\, \Omega ^4+7803\, \Omega ^3+9378\, \Omega ^2+5454 \,\Omega +1215\right)}{180 (2\, \Omega +3)^5 (2 \,\Omega +5) (2\, \Omega +7)}.
\notag
\end{align}
The expansion of $\kappa$ in (\ref{3.5}), may be turned into an expansion of $\Omega$ according to 
\begin{align}
\la{3.11}
\,\Omega =  & \frac{t-1}{2}+
\frac{(t-1) (t+1)^2 \phi ^2}{16 t (t+2)} +\notag \\
&\frac{(t-1) (t+1)^2 \left(5 \,t^5+40 \, t^4+97\, t^3+68 \, t^2+18 \, t+24\right) \phi ^4}{768 t^3 (t+2)^3 (t+4)} + \\
& \frac{(t-1) (t+1)^2}{92160 t^5 (t+2)^5 (t+4)^2 (t+6)} \left(61\, t^{11}+1342 \, t^{10}+12170\, t^9+58906\,t^8+163733\,t^7+
\right. \notag \\ 
& \qquad \left.259846 \, t^6+222676\, t^5+107946\, t^4+65040\, t^3+57960\, t^2+37440\, t+17280\right) \phi ^6 + \dots \notag
\end{align}
where $t=\sqrt{\kappa+1}$. The general structure of the higher order corrections is 
\be
\la{3.12}
\Omega=\sum_{n=1}^{\infty}\mathcal{P}_n(t)  \mathcal{D}_n(t) \phi^{2n},
\ee
where $\mc D_{n}(t)$ is the rational function 
\be
\label{3.13}
\begin{split}
\mathcal{D}_n(t) = \frac{(t+2)^2 (t-1) (t+1)^2}{\left[ 2^{2 (n+1)-1} t^{n-2} \Gamma (2 (n+1)-1)\right] \prod _{j=1}^{n+1} t (t+2) (2 j+t)^{-j+n+1}},
\end{split}
\ee
and $\mc P_{n}(t)$ are polynomials. The first cases are collected in App.~(\ref{app:list}). They have an increasing 
complexity, but may be generated quite easily by the above procedure.

\subsection{A cross-check at large $\kappa$}
\label{subsec:large}

The expansion (\ref{3.11}) may be checked at weak coupling, {\em i.e.} in the limit 
$\kappa\to 0$, using the six-loop expressions
derived in \cite{Henn:2012qz}. At strong coupling, we can evaluate systematically the perturbative 
expansion of the (ladder) cusp anomaly for any $\phi$. This can be achieved by the same strategy
described in the previous section, {\em i.e.} by an Ansatz similar to (\ref{3.5}). We begin 
by multiplying the \schr equation 
 by $\mu^{2}/\kappa$ with $\mu^{2} = 4\,\sqrt{\kappa}(\cos\phi+1)$. After a rescaling 
 $x = \mu\,X/\sqrt{\kappa}$, we obtain 
\be
\label{3.14}
\begin{split}
& \bigg(-\frac{d^{2}}{dX^{2}}+\mc V(X, \phi) \bigg)\,\psi(X) = E\,\psi(X), \\ 
& \mc V(X,\phi) = -\frac{\sqrt\kappa}{2}+X^{2}+\frac{\cos\phi-5}{3\,\sqrt\kappa}\,X^{4}
+\frac{123-56\,\cos\phi+\cos(2\phi)}{45\,\kappa}\,X^{6}+\dots, \\
E &= -\frac{\cos\phi+1}{\sqrt\kappa}\,\Omega^{2}.
\end{split}
\ee
The perturbative correction to the ground state energy $E$ of the (formal) problem 
\be
\la{3.15}
-\psi''(X)+\bigg(X^{2}+\sum_{n=2}^{\infty} \varepsilon^{n-1}\,c_{2n}\,X^{2n}\bigg)\,\psi(X) = E\,\psi(X),
\ee
can be found efficiently by the Ansatz
\be
\la{3.16}
\psi(X) = e^{-\frac{X^{2}}{2}}\bigg(1+\sum_{n=1}^{\infty} p_{n}(X)\,\varepsilon^{n}\bigg),
\ee
where $p_{n}(X)$ is an even polynomial with degree $4n$ and without constant term. The first three cases are 
explicitly
\begin{align}
\la{3.17}
p_{1}(X) &= -\frac{3 c_4 X^2}{8}-\frac{c_4 X^4}{8}, \\
p_{2}(X) &= \frac{3}{32} \left(7 c_4^2-10 c_6\right) X^2+\frac{1}{128} \left(31 c_4^2-40 c_6\right)
   X^4+\frac{1}{192} \left(13 c_4^2-16 c_6\right) X^6+\frac{1}{128} c_4^2
   X^8, \notag \\
p_{3}(X) &= -\frac{3}{128} \left(111 c_4^3-240 c_6 c_4+140 c_8\right) X^2+\frac{1}{256} \left(-243
   c_4^3+510 c_6 c_4-280 c_8\right) X^4 \notag \\
   &+\frac{1}{3072}\,\left(-813 c_4^3+1680 c_6 c_4-896 c_8\right)\,
   X^6+\frac{1}{1024}\,\left(-47 c_4^3+104 c_6 c_4-64 c_8\right) \,X^8 \notag \\
   &-\frac{c_{4}}{3072}\,
   \left(17 c_4^2-32 c_6\right)\,X^{10}-\frac{c_4^3
   }{3072}\,X^{12},\notag
\end{align}
and lead to the perturbed energy
\be
\la{3.18}
E = 1+\frac{3\,c_{4}}{4}\,\varepsilon-\frac{3}{16}\,(7\,c_{4}^{2}-10\,c_{6})\,\varepsilon^{2}
+\frac{3}{64}\,(111\,c_{4}^{3}-240\,c_{4}\,c_{6}+140\,c_{8})\,\varepsilon^{3}+\cdots.
\ee
The associated expansion of the cusp anomalous dimension is therefore
\be
\la{3.19}
\begin{split}
\Gamma^{\text{lad}} &= \frac{\sqrt{\kappa}}{2\,\cos\frac{\phi}{2}}\bigg[
1-\frac{1}{\sqrt\kappa}+\frac{3-\cos\phi}{4\,\kappa}+\frac{1}{\kappa^{3/2}}\,
\frac{-21+20\,\cos\phi+\cos(2\phi)}{32}\\
&+\frac{1}{\kappa^{2}}
\,\frac{290-317\,\cos\phi-2\,\cos(2\phi)-3\,\cos(3\phi)}{256}+\cdots\bigg],
\end{split}
\ee
in full agreement with what one finds by expanding  (\ref{3.11}) at large $t$.

\subsection{Numerical analysis}
\label{subsec:num}

\begin{figure}[h!]
\centering
\includegraphics[width=0.49\textwidth]{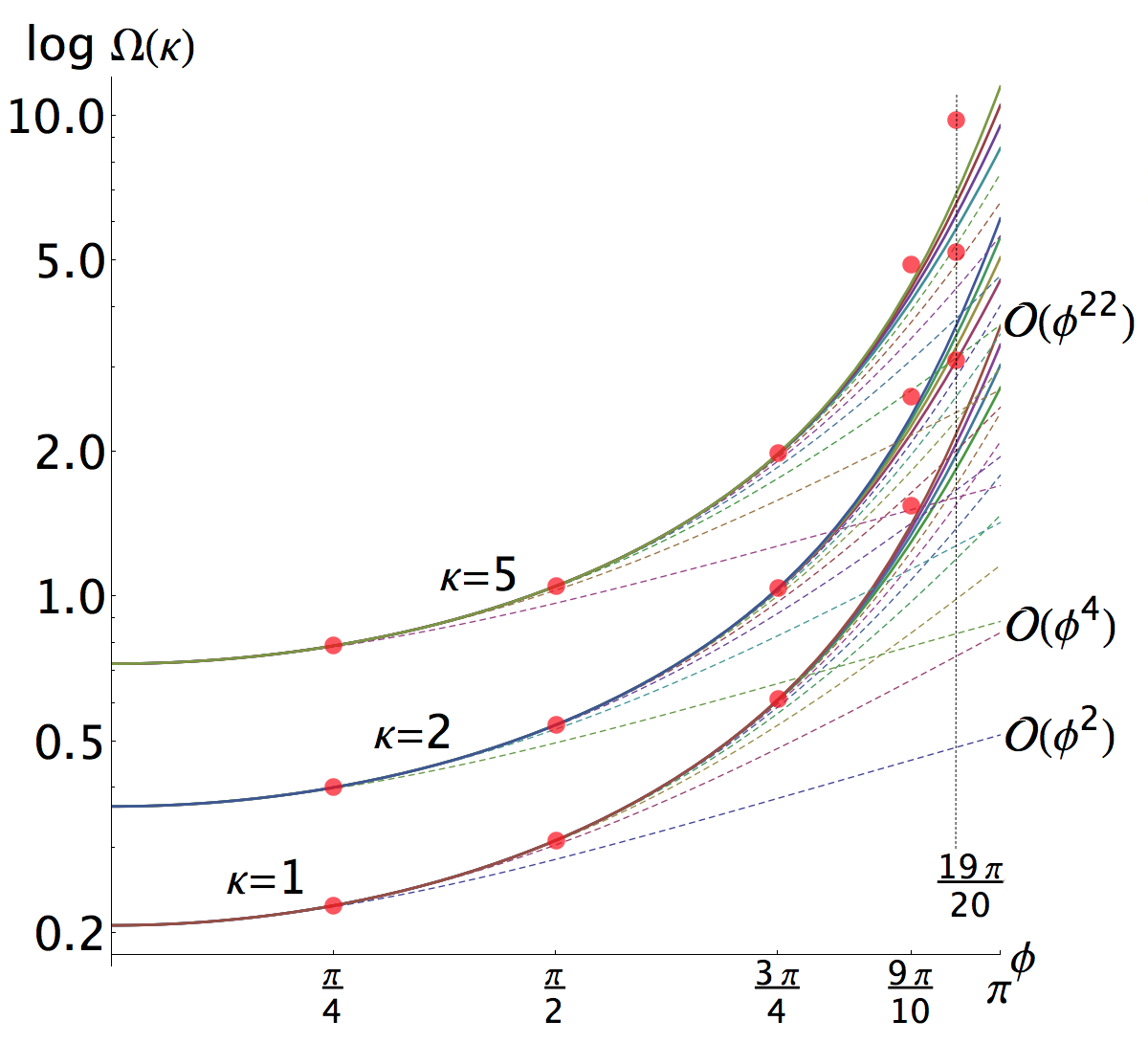}
\includegraphics[width=0.49\textwidth]{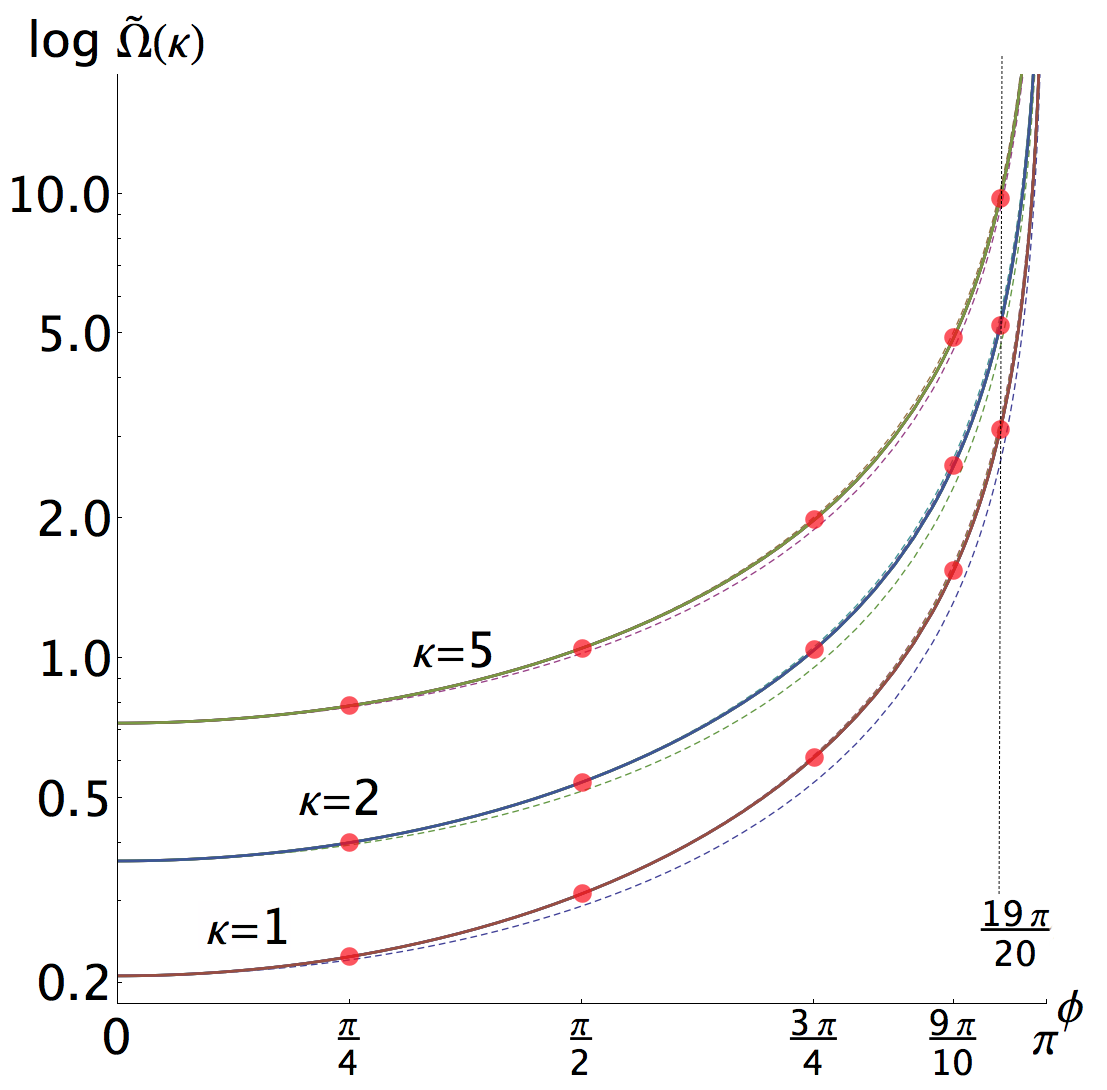}
\caption{The plots show the convergence of the perturbative expansion of $\Omega$ at small $\phi$ 
for $\kappa = 1,2,5$. The many thin dashed lines show results with 
lower values of $N$ in (\ref{3.20}) (left panel) and (\ref{3.21}) (right panel). 
The four red dots are the (numerical) exact values of $\Omega$ obtained solving numerically the 
\schr equation at  $\phi = \tfrac{\pi}{4}, \tfrac{\pi}{2}, \tfrac{3\,\pi}{4}$, $\tfrac{9\,\pi}{10}$, and 
$\tfrac{19\,\pi}{20}$, see also   Tab.~(\ref{tab1}).}
\label{2}
\end{figure}

In this section, we compare the exact numerical solution of the \schr problem (\ref{1.3}) with the small $\phi$
expansion of its ground state energy, see  (\ref{3.12}). We shall consider three reference values $\kappa=1,2,5$ and explore convergence with respect to 
 $\phi$ in the physical  interval $0\le \phi<\pi$. The first comparison is with the naive partial sums 
 of (\ref{3.12}),  ( recall that $\Gamma^{\rm lad} = -\Omega$ ) 
 \be
 \la{3.20}
 \Omega_{N} = \sum_{n=1}^{N}\,  \mc P_{n}\,\mc D_{n}\,\phi^{2n}.
 \ee
 This is shown in the left panel of Fig.~(\ref{2}). 
 One sees that including terms up to $N=20$, there is convergence to
 the exact numerical value at least for  $\phi$ not too close to $\pi$. As $\phi\to \pi$, convergence slows down and 
 the series expansion cannot provide an accurate estimate of the correct (ladder) cusp anomaly. This is clearly 
 illustrated by the points at $\phi = \tfrac{19\pi}{20}$.
 However, the strong coupling expansion (\ref{3.19}) suggests that the singularity at $\phi=\pi$ is simply due to an overall
 factor $1/\cos(\phi/2)$. 
 From the physical point of view, the $\phi\to \pi$ limit is a flat space limit where that singularity 
 is nothing but the overall scale $1/r$ in the quark-antiquark at distance $r \sim 2\,\cos(\frac{\phi}{2})$ \cite{Correa:2012nk}.
 Hence, we also compare the numerical values of the cusp anomaly with the improved summation 
 \be
 \la{3.21}
\widetilde  \Omega_{N} = \frac{1}{\cos\frac{\phi}{2}}\left[
\cos\frac{\phi}{2}\sum_{n=1}^{N}\, \mc P_{n}\,\mc D_{n}\,\phi^{2n}\right]_{\text{truncated at $\phi^{2N}$}}.
 \ee
 The right panel of Fig.~(\ref{2}) shows that this is a major improvement. The convergence in $N$ is 
 now greatly increased  and
  accurate results are obtained even quite near the singular point $\phi=\pi$. This is illustrated in a more quantitative way
 in Tab.~(\ref{tab1}) where we collect  some reference numerical values  shown in  Fig.~(\ref{2}). In all cases, the 
 relative accuracy of (\ref{3.21}) is well below the $10^{-3}$ level.
\begin{table}[h!]
\be
\notag
\renewcommand{\arraystretch}{1.4}
\begin{array}{|cc|c|c|c|}
\hline
\kappa & \phi & \Omega\,\, \text{exact} & \Omega_{N=20} & \widetilde\Omega_{N=20} \\
\hline
1 & \frac{\pi}{4} 	& 0.2277386205  &\; \blue{0.2277386205}  &\; \blue{0.2277386205}   \;\; \\
1 & \frac{\pi}{2} 	& 0.3112053734 &\; \blue{0.3112053}487  &\; \blue{0.311205373}3   \;\;\\
1 & \frac{3\pi}{4} 	& 0.6121866441	&\; \blue{0.61}14749855  &\; \blue{0.61218}58951   \;\; \\
1 & \frac{9\pi}{10} 	& 1.546211922 	&\; \blue{1}.4159044310  &\; \blue{1.546}0568021   \;\;\\
1 & \frac{19\pi}{20} 	& 3.101434441	&\;       2.17175679664	 &\; \blue{3.10}02145584   \;\;\\
\hline
2 & \frac{\pi}{4} 	& 0.4008519359 	&\;  \blue{0.400851935}8 &\;  \blue{0.4008519359}  \;\; \\
2 & \frac{\pi}{2} 	& 0.5409421028	&\;  \blue{0.540942}0614 &\;  \blue{0.540942102}7  \;\; \\
2 & \frac{3\pi}{4} 	& 1.042364136	&\;  \blue{1.04}11714765 &\;  \blue{1.04236}35475  \;\; \\
2 & \frac{9\pi}{10} 	& 2.599160526   &\;  \blue{2}.3808426079 & \; \blue{2.599}0745817  \;\; \\
2 & \frac{19\pi}{20} 	& 5.199351482	&\;        3.6420358922  & \; \blue{5.19}88023024  \;\; \\
\hline
5 & \frac{\pi}{4} 	& 0.7898806509 	&\; \blue{0.789880650}8  &\;  \blue{0.789880650}8  \;\;\\
5 & \frac{\pi}{2} 	& 1.051108721 	&\; \blue{1.051108}6429  &\;  \blue{1.05110872}06  \;\;\\
5 & \frac{3\pi}{4} 	& 1.985138624 	&\; \blue{1.98}28960288  &\;  \blue{1.985138}9346  \;\;\\
5 & \frac{9\pi}{10} 	& 4.899715022 	&\; \blue{4}.489210222 	 &\;  \blue{4.8997}700298  \;\;\\
5 & \frac{19\pi}{20} 	& 9.783226208	&\;       6.8549556957	 &\;  \blue{9.783}6202372  \;\;\\
\hline
\end{array}
\ee
\caption{This table compares the exact values of $\Omega$ with the approximations in (\ref{3.20}) and (\ref{3.21}) for
$N=20$. The comparison is done at $\kappa = 1,2,5$ and various $\phi$. The
first column contains the (numerical) exact value corresponding to the red dots in Fig.~(\ref{2}).
Digits highlighted in blue are in agreement with the exact solution.}
\label{tab1}
\end{table}

\section{Conclusions}

In this paper we have considered various properties of the perturbative expansion of the cusp anomalous dimension
in $\mc N=4$ SYM in the (leading order) ladder approximation. We have presented simple algorithms for the 
higher-order evaluation of  the (i) small $x = e^{i\,\phi}$ and  (ii) small $\phi$ corrections.
In the former case, we showed that all the logarithmic corrections
are captured by a Wood-Saxon type solvable problem. Besides, we have shown how to generate all such corrections
at higher-orders by compact generating functions that encode them and are naturally organised in increasing 
transcendentality degree. Our approach explains the remarkable regularities observed in the past. 
In the small $\phi$ regime, we showed that is possible to work out the quantum mechanical perturbation
expansion bypassing the  Rayleigh-\schr scheme. This is due to the simple structure of the perturbed wave-function
associated with the ground state. Our remark leads to a quite efficient algorithm.
The associated long expansion in powers of $\phi$ has been shown to provide, after some educated manipulations, 
an accurate  representation of the ladder cusp anomaly in the whole range of couplings and angles.
We believe that it would be very interesting to look at our results from the perspective of the Quantum Spectral
Curve by extending the analysis of \cite{Gromov:2016rrp} to the case of a generic cusp  angle $\phi$. 
Our new results could certainly be  useful  as a non-trivial check of that method.

\appendix

\section{Complete expression of the eight-loops term $\Omega_{8}$}
\label{app:eight}

\begin{align}
\label{A.1}
 &\Omega_8(x) \,=\,
\frac{13955497984}{638512875} \, \log^{15}x +\frac{1744437248}{18243225} \,\pi ^2 \, \log^{13}x +\frac{4083712}{13365} \,\zeta _3 \, \log^{12}x
 \notag \\&\quad  
+ \frac{371067328}{2338875} \,\pi ^4 \, \log^{11}x +\left(\frac{8167424 \,\pi ^2 \,\zeta _3}{8505}+\frac{17188736 \,\zeta_5}{14175}\right) \log^{10}x
 \notag \\&\quad  
+ \left(\frac{1356032 \,\zeta _3^2}{945}+\frac{110648288 \,\pi ^6}{893025}\right)\log^9 x+\left(\frac{4964096 \,\pi ^4 \,\zeta _3}{4725}+\frac{12736 \,\pi ^2 \,\zeta _5}{5}
\right. \notag \\&\quad \left.  
+ \frac{378176 \,\zeta _7}{105}\right) \log^8 x+
\left(\frac{2712064}{945} \,\pi ^2\,\zeta _3^2+\frac{2722816 \,\zeta _5 \,\zeta _3}{315}+\frac{205490336 \,\pi ^8}{4465125}\right)\, \log^7 x
 \notag \\&\quad  
+\left(\frac{68608 \,\zeta _3^3}{27}+\frac{20108288 \,\pi ^6 \,\zeta_3}{42525}+\frac{124928 \,\pi ^4 \,\zeta _5}{75}+\frac{205568 \,\pi ^2 \,\zeta_7}{45}+ \frac{1036864 \,\zeta _9}{135}\right) \, \log^6 x
 \notag \\&\quad  
+\left(\frac{5120}{3} \,\pi ^4 \,\zeta _3^2+\frac{49152}{5} \,\pi ^2 \,\zeta _5 \,\zeta _3+17152 \,\zeta _7 \,\zeta _3+8576 \,\zeta_5^2+ \frac{17034488 \,\pi ^{10}}{2338875}\right) \, \log^5 x
 \notag \\&\quad  
+\left(\frac{68608}{27} \,\pi ^2 \,\zeta _3^3+\frac{42752}{3} \,\zeta _5 \,\zeta _3^2+ \frac{682912 \,\pi ^8 \,\zeta_3}{8505}+\frac{1050944 \,\pi ^6 \,\zeta _5}{2835}+\frac{67664 \,\pi ^4 \,\zeta_7}{45}
\right. \notag \\&\quad \left.
+ \frac{135920 \,\pi ^2 \,\zeta _9}{27}+\frac{33424 \,\zeta _{11}}{3}\right) \, \log^4 x+ \left(\frac{11776 \,\zeta _3^4}{9}+\frac{887552 \,\pi ^6 \,\zeta _3^2}{2835}+\frac{118144}{45} \,\pi ^4 \,\zeta _5 \,\zeta _3
\right. \notag \\&\quad \left. 
+\frac{26752}{3} \,\pi ^2 \,\zeta _7\,\zeta _3+\frac{179072 \,\zeta _9 \,\zeta _3}{9}+\frac{13376}{3} \,\pi ^2 \,\zeta _5^2+19584 \,\zeta _5 \,\zeta _7+\frac{18213284 \,\pi ^{12}}{49116375}\right) \, \log^3 x
 \notag \\&\quad 
+ \left(\frac{23168}{45}\,\pi ^4 \,\zeta _3^3+5504 \,\pi ^2 \,\zeta _5 \,\zeta _3^2+12672 \,\zeta _7 \,\zeta _3^2+ 12672 \,\zeta_5^2 \,\zeta _3+\frac{190016 \,\pi ^{10} \,\zeta _3}{51975}+ \frac{6704 \,\pi ^8 \,\zeta_5}{315}
\right. \notag \\&\quad \left. 
+\frac{36832 \,\pi ^6 \,\zeta _7}{315}+\frac{27352 \,\pi ^4 \,\zeta _9}{45}+2832 \,\pi ^2\,\zeta _{11}+10056 \,\zeta _{13}\right) \, \log^2 x
 \notag \\&\quad  
+ \left(\frac{2944}{9} \,\pi ^2 \,\zeta_3^4+\frac{9728}{3} \,\zeta _5 \,\zeta _3^3+\frac{160096 \,\pi ^8 \,\zeta_3^2}{14175}+\frac{17024}{135} \,\pi ^6 \,\zeta _5 \,\zeta _3+\frac{3232}{5} \,\pi ^4 \,\zeta _7 \,\zeta _3
\right.\notag \\&\quad \left.   
+\frac{26624}{9} \,\pi ^2 \,\zeta _9 \,\zeta _3+10336 \,\zeta _{11} \,\zeta_3+\frac{1616}{5} \,\pi ^4 \,\zeta _5^2+4752 \,\zeta _7^2+2880 \,\pi ^2 \,\zeta _5 \,\zeta_7
\right. \notag \\&\quad \left.  
+\frac{29344 \,\zeta _5 \,\zeta _9}{3}+\frac{1859138 \,\pi ^{14}}{638512875}\right) \log x+\frac{256 \,\zeta _3^5}{5}+ \frac{10112}{945} \,\pi ^6 \,\zeta _3^3+864 \,\zeta _5^3+768 \,\pi ^2 \,\zeta _3 \,\zeta _5^2
\notag \\&\quad  
+\frac{87376 \,\pi ^{12} \,\zeta _3}{6081075}+\frac{2528}{15} \,\pi ^4 \,\zeta _3^2 \,\zeta _5+ \frac{5528 \,\pi ^{10} \,\zeta _5}{51975}+768 \,\pi ^2 \,\zeta _3^2 \,\zeta_7+5184 \,\zeta _3 \,\zeta _5 \,\zeta _7+ \frac{248 \,\pi ^8 \,\zeta _7}{315}
\notag \\&\quad  
+2656 \,\zeta _3^2 \,\zeta_9+\frac{1156 \,\pi ^6 \,\zeta _9}{189}+\frac{248 \,\pi ^4 \,\zeta_{11}}{5}+420 \,\pi ^2 \,\zeta_{13}+\frac{21844 \,\zeta _{15}}{5}+\mc O(x).
\end{align}

\section{Convergence of the expansion (\ref{2.13})}
\label{app:conv}

The expansion (\ref{2.13}) solves (\ref{2.12}). The dependence on powers of $x$ and $\log x$ is 
clearly trivial. Indeed, introducing  the variables
\be
\la{B.1}
\kappa = \frac{\overline \kappa}{x\,\log^{2} x}, \qquad \omega_{0} = \frac{\overline \omega_{0}}{\log x},
\ee
we can rewrite (\ref{2.12}) as 
\be
\la{B.2}
\sqrt{\overline \kappa-\overline\omega_{0}^{2}}\,
\tan\bigg(\frac{1}{2}\sqrt{\overline \kappa-\overline\omega_{0}^{2}}\bigg)+\overline\omega_{0} = 0.
\ee
This equation admits a solution $\overline\omega_{0}(\overline \kappa)$ that is analytic in a disc of radius $R$.
This convergence radius is determined by a branch point at $\overline\kappa<0$ where a pair of real 
roots of (\ref{B.2}) coalesce into a pair of complex conjugate roots. Setting $\overline \kappa = -R>0$, 
we determine $R$ by eliminating $w$ in the two equations
\be
\la{B.3}
F(R, w) = w-\sqrt{R+w^{2}}\,\tanh\bigg(\frac{1}{2}\sqrt{R+w^{2}}\bigg)=0, 
\qquad 
\frac{\partial}{\partial w}\, F(R, w) = 0.
\ee
This gives $w=2$. Then, $R$ is found as the unique positive root of 
\be
\la{B.4}
\sqrt{R+4}\,\tanh\bigg(\frac{1}{2}\,\sqrt{R+4}\bigg) = 2,
\ee
that is $R=1.756915\dots$ and (\ref{2.13}) converges for $|\kappa|<R/(x\,\log^{2}x)$.

\section{Small $x$ limit as  perturbation around a finite depth well}
\label{app:qm}

The potential in (\ref{2.5}) can also be split in the following way, where we have again 
 rescaled $w$ by $\log x$ introducing $w = \tau\, \log x$ for $\tau>0$,
\be
\la{C.1}
-\frac{1/x}{\cosh(\tau\,\log x)+\frac{1}{2}(x+x^{-1})} = -2\,\Theta(1-\tau) + 
\frac{2\,\text{sign}(1-\tau)}{1+e^{-\log x\,|\tau-1|}}+\dots\, .
\ee
where $\Theta(\tau)$ is the Heaviside step function. In other words, the potential looks like a Fermi-Dirac distribution.
The  unperturbed shape is a finite depth well, while the correction -- the second term in (\ref{C.1}) -- captures
its deviation in the small strip $\tau-1\sim 1/\log x$.

\medskip
The solution of the Schr\"odinger equation for the finite depth well is elementary and reads, for $\tau>0$, 
\be
\la{C.2}
\psi(\tau) = \begin{cases}
A\,\cos\bigg(\log x\,\frac{\sqrt{\kappa\, x-\Omega^{2}}}{2}\, \tau\bigg), & 0<\tau<1, \\
B\,\exp\bigg(\frac{\Omega\,\log x}{2}\,\tau\bigg), & \tau>1. 
\end{cases}
\ee
At $\tau=1$, continuity fixes the ratio $A/B$, while continuity of $\psi'$ determines the relation between 
$\Omega$ and $\kappa$ to be precisely (\ref{2.12}).

\medskip
This first approximation may be improved by taking into account the correction in (\ref{C.1}). Treating it at first 
order in perturbation theory amounts to compute its integral times $|\psi(\tau)|^{2}$. But this is accomplished as 
follows (we omit a trivial factor 2)
\be
\la{C.3}
\begin{split}
& \int_{0}^{\infty} d\tau\, \frac{\text{sign}(1-\tau)}{1+e^{-\log x\,|\tau-1|}}\, F(\tau) = 
\int_{0}^{1} d\tau\, \frac{F(\tau)}{1+e^{-\log x\,(1-\tau)}}
-\int_{1}^{\infty} d\tau\, \frac{F(\tau)}{1+e^{-\log x\,(\tau-1)}}\\
&= -\frac{1}{\log x}\, \int_{0}^{-\log x}d\mu\,\frac{F(1+\frac{\mu}{\log x})}{1+e^{\mu}}
+\frac{1}{\log x}\,\int_{0}^{\infty}d\mu\,\frac{F(1-\frac{\mu}{\log x})}{1+e^{\mu}}
\end{split}
\ee
Up to corrections $\mc O(x)$ -- from the upper integration limit in the first integral -- we can expand at small $x$ 
and obtain 
\be
\la{C.4}
\begin{split}
& \int_{0}^{\infty} d\tau\, \frac{\text{sign}(1-\tau)}{1+e^{-\log x\,|\tau-1|}}\, F(\tau) = 
\int_{0}^{\infty}d\mu\,\frac{1}{1+e^{\mu}} \\
& \bigg[
\frac{F(1^{-})-F(1^{+})}{\log x}
-\mu\,\frac{F'(1^{-})+F'(1^{+})}{\log^{2} x}
+\mu^{2}\,\frac{F''(1^{-})-F''(1^{+})}{2\,\log^{3} x}+\cdots
\bigg] .
\end{split}
\ee
In our application $F(\tau) = |\psi(\tau)|^{2}$, and $F'(\tau)$ are  continuous at $\tau=1$, so 
\be
\la{C.5}
\begin{split}
& \int_{0}^{\infty} d\tau\, \frac{\text{sign}(1-\tau)}{1+e^{-\log x\,|\tau-1|}}\, F(\tau) =  -\frac{\zeta_{2}}{\log^{2}x}\,F'(1)+\frac{3}{4}\frac{\zeta_{3}}{\log^{3} x}\,
[F''(1^{-})-F''(1^{+})]+\cdots\ .
\end{split}
\ee
Collecting all pieces, we obtain the simple formula 
\be
\label{C.6}
\Omega^{2}_{\text{NNLO}} = \Omega^{2}+4\,\kappa\,x\,(\delta E_{1}+\delta E_{2}^{+}+\delta E_{2}^{-}),
\ee
where $\Omega$ is the leading order, {\em i.e.} the solution of (\ref{2.12}), and 
\be
\label{C.7}
\begin{split}
\delta E_{1} &= \frac{\zeta_{2}}{2\,\mc N}\,\frac{\psi(1)\,\psi'(1)}{\log^{2}x}, \qquad
\delta E_{2}^{\pm} = \mp\,\frac{\zeta_{3}}{2\,\mc N}\frac{\psi(1)\,\psi''(1^{\pm})+\psi'(1)^{2}}{\log^{3}x}, \\
\mc N &= \int_{0}^{\infty} d\tau\,|\psi(\tau)|^{2}.
\end{split}
\ee
After some simplification, it is possible to show that  (\ref{C.6}) is indeed equivalent to the first three
terms of (\ref{2.15}). In this approach, the appearance of the simple transcendental zeta values if simply due to the elementary integral
\be
\la{C.8}
\int_{0}^{\infty}d\mu\, \frac{\mu^{n}}{1+e^{\mu}} = (1-2^{-n})\,n!\,\zeta_{n+1}.
\ee
The leading perturbative calculation in (\ref{C.6}), already captures the exact expansion at NNLO.
The second order perturbation with respect to the second term in (\ref{C.1}) will give a $\zeta_{2}^{2}$
contribution that mixes with a genuine $\zeta_{4}$ term from dots in (\ref{C.5}). Despite its semplicity, this 
method cannot be extended in a simple way to higher orders because of the complexity of the expressions 
for the higher order correction to the energy. These involve the full spectrum as well as infinite sums over
the unperturbed wave-functions. However, as we discussed in the main text, this complexity is only apparent
due to the solvability of the potential in (\ref{2.5}).

\section{List of higher order corrections for small $\phi$ up to $\mc O(\phi^{14})$}
\la{app:list}

{\footnotesize
\begin{align}
\label{D.1}
\mathcal{P}_1(t) = & 1,
\notag \\
\mathcal{P}_2(t) = & 5 t^5+40 t^4+97 t^3+68 t^2+18 t+24,
\notag \\
\mathcal{P}_3(t) = & 61 t^{11}+1342 t^{10}+12170 t^9+58906 t^8+163733 t^7+259846 t^6+222676 t^5+107946 t^4+65040 t^3+
\notag \\&\qquad
57960 t^2+37440 t+17280,
\notag \\
\mathcal{P}_4(t) = & 1385 t^{18}+60940 t^{17}+1195947 t^{16}+13843810 t^{15}+105276563 t^{14}+553998174 t^{13}+2067538027 t^{12}+
\notag \\&\qquad
5507959178 t^{11}+10383911640 t^{10}+13546139774 t^9+11889676294 t^8+7154803452 t^7+3839607792 t^6+
\notag \\&\qquad
2842367472 t^5+2492475552 t^4+2180304000 t^3+1653926400 t^2+861235200 t+232243200,
\notag \\   
\mathcal{P}_5(t) = & 50521 t^{26}+3839596 t^{25}+134966852 t^{24}+2915424324 t^{23}+43365024790 t^{22}+471649242590 t^{21}+
\notag \\&\qquad
3887215754528 t^{20}+24822724030896 t^{19}+124499947421625 t^{18}+494084918575016 t^{17}+
\notag \\&\qquad
1554906212199248 t^{16}+3870173006254276 t^{15}+7561022864208816 t^{14}+11448441114126974 t^{13}+
\notag \\&\qquad
13212006774954052 t^{12}+11457199366013944 t^{11}+7582649551861488 t^{10}+4335757273241184 t^9+
\notag \\&\qquad
2789712226451520 t^8+2204964605554560 t^7+1927177447253760 t^6+1807025647472640 t^5+
\notag \\&\qquad   
1665731489587200 t^4+1315285674393600 t^3+781170440601600 t^2+300585870950400 t+56184274944000,
\notag \\
\mathcal{P}_6(t) = & 2702765 t^{35}+324331800 t^{34}+18377450617 t^{33}+654386062352 t^{32}+16440800222532 t^{31}+
\notag \\&\qquad
310199647784458 t^{30}+4567661514832408 t^{29}+53851031228968518 t^{28}+517442799401984195 t^{27}+
\notag \\&\qquad
4103665112187249680 t^{26}+27101585423951848065 t^{25}+149963124703098879052 t^{24}+
\notag \\&\qquad
697919503178137618918 t^{23}+2736906706749509241786 t^{22}+9043685297188043209022 t^{21}+
\notag \\&\qquad
25133316635666679058526 t^{20}+58519478658626463125526 t^{19}+113469650417310829074772 t^{18}+
\notag \\&\qquad
181699082261549829944864 t^{17}+237759675035517534720960 t^{16}+251372784157888368583264 t^{15}+
\notag \\&\qquad
213305246564456712530688 t^{14}+147516618960066395589504 t^{13}+89936991802153035643392 t^{12}+
\notag \\&\qquad
57126447910320186240000 t^{11}+42292252096794621398016 t^{10}+34454605595328501073920 t^9+
\notag \\&\qquad
29634415605623306649600 t^8+27539912981290335436800 t^7+27046152016870440960000 t^6+
\notag \\&\qquad
25539521870164171161600 t^5+20782297347901725081600 t^4+13307907933087989760000 t^3+
\notag \\&\qquad
6151734205648522444800 t^2+1813380285434167296000 t+256308167552532480000,
 \notag \\
\mathcal{P}_7(t) = & 199360981 t^{45}+35486254618 t^{44}+3022212791470 t^{43}+164023100493262 t^{42}+
\notag \\&\qquad
6374655375179395 t^{41}+189009201830969872 t^{40}+4447885774562687212 t^{39}+
\notag \\&\qquad
85331210261611492012 t^{38}+1360394617537100985643 t^{37}+18280296531547559838528 t^{36}+
\notag \\&\qquad
209279455515929250214386 t^{35}+2058124179013507585837284 t^{34}+17497246756285714164347853 t^{33}+
\notag \\&\qquad
129217605320007302424285456 t^{32}+831946420045744026597574272 t^{31}+4681727333630757032920147884 t^{30}+
\notag \\&\qquad
23065893592425253503899606208 t^{29}+99572159399302861728248648046 t^{28}+
\notag \\&\qquad
376624515246828447449840837492 t^{27}+ 1247186722497447982915317705494 t^{26}+
\notag \\&\qquad
3609860729195291609521980084752 t^{25}+9109060373443498758351292201688 t^{24}+
\notag \\&\qquad 
19967717605312852001306027530592 t^{23}+37844932785162123592573594988192 t^{22}+
\notag \\&\qquad 
61648377450275703706508745675008 t^{21}+85691979165642836742445115368064 t^{20}+
\notag \\&\qquad  
100823918241913794022079588633600 t^{19}+99672905219977016061123259740672 t^{18}+
\notag \\&\qquad 
82610273059021399617505652981760 t^{17}+58285708981399017149841984227328 t^{16}+
\notag \\&\qquad 
37103058887642908565106794766336 t^{15}+23995413510276589877794905907200 t^{14}+
\notag \\&\qquad 
17508929989269158330481642700800 t^{13}+14136689270785463628330509107200 t^{12}+
\notag \\&\qquad 
11789812055828945185925682954240 t^{11}+10012775577450271647121775001600 t^{10}+
\notag \\&\qquad    
8962503171944994703109888409600 t^9+8692238054012201609624498995200 t^8+ 
\notag \\&\qquad   
8777659166798668516896119193600 t^7+8399023861484287111423485542400 t^6+
\notag \\&\qquad   
6963039926089402427394490368000 t^5+4682955977911720432782802944000 t^4+
\notag \\&\qquad   
2415338217244404925718121676800 t^3+891894652452125893831163904000 t^2+
\notag \\&\qquad   
209764769616392065426391040000 t+23644982082363034750156800000
\end{align}
}

\bibliography{QQbar-Biblio}
\bibliographystyle{JHEP}

\end{document}